\documentclass[aps,pra, twocolumn,longbibliography]{revtex4-1}
\usepackage{graphicx}
\usepackage{amsmath}
\usepackage{amssymb}
\usepackage{braket}
\usepackage{hyperref}
\usepackage{subfigure}
\usepackage[usenames,dvipsnames]{color}

\usepackage{graphicx}
\usepackage{subfigure}
\usepackage[mathscr]{euscript}

\newcommand{\be}{\begin{equation}}
\newcommand{\ee}{\end{equation}}
\newcommand{\ba}{\begin{eqnarray}}
\newcommand{\ea}{\end{eqnarray}}

\newcommand{\non}{\nonumber}
\begin{document}

\title{Quasi-localization dynamics in a Fibonacci quantum rotor}

\author{Sourav Bhattacharjee}
\email{sourav.offc@gmail.com}
\author{Souvik Bandyopadhyay}
\author{Amit Dutta}
\affiliation{Department of Physics, Indian Institute of Technology Kanpur, Kanpur 208016, India}

\begin{abstract}
We analyze the dynamics of a quantum kicked rotor (QKR) driven with a binary Fibonacci sequence of two distinct drive amplitudes. While the dynamics at low drive frequencies is found to be diffusive, a long-lived pre-ergodic regime emerges in the other limit. Further, the dynamics in this pre-ergodic regime  can be associated with the onset of a dynamical quasi-localization, similar to the dynamical localization observed in a regular QKR.
We establish that this peculiar behavior arises due to the presence of localized eigenstates of an approximately conserved effective Hamiltonian, which drives the evolution at Fibonacci instants. However, the effective Hamiltonian picture does not persist indefinitely and the dynamics eventually becomes ergodic after asymptotically long times. 
\end{abstract}

\maketitle

\section{Introduction}\label{sec_intro}	
The quantum kicked rotor (QKR) \cite{chirikov79, izrailev90} is central to the understanding the basics of quantum chaos and has been subjected to a plethora of analytical \cite{casati79,fishman82, fishman84, chang86, fishman87, fishman89} as well as experimental \cite{galvez88, bayfield89, moore94, moore95, ammann98, bitter16, sarkar17} investigations over the years \cite{stockmann99, haake18, anatoli16, santhanam21}.  
In contrast to the classical rotor which shows a transition from a regular to the chaotic phases, the QKR seemingly exhibits a non-ergodic behavior. Using Floquet theory \cite{floquet83, shirley65, bukov15, eckardt17}, it is found that the eigen-states of the effective Floquet Hamiltonian are exponentially localized in the angular momentum space. This, together with quantum interference affects, lead to dynamical localization in periodic QKRs. Remarkably, the non-ergodicity in the dynamics is manifested for any finite amplitude and frequency of the kicks (see Appendix~\ref{app_regular} for a brief recapitulation).\\

The \textit{dynamical localization} observed in the QKR has far reaching implications. It was realised that this localization can be exactly mapped to the spatial Anderson localization problem \cite{anderson58} in one-dimension \cite{fishman82, fishman84}. However, the Anderson problem in more than one dimension is known to exhibit a localization-delocalization transition \cite{abrahams79}. In the case of the QKR, similar transitions were also found, albeit in a slightly modified version in which the rotor is driven with three incommensurate frequencies. In this case, the rotor Hamiltonian with time dependent kick amplitude is first mapped to a three dimensional rotor \cite{3d_rotor_note} with time-independent kick amplitudes at equal time intervals \cite{casati89}. The temporal evolution at stroboscopic intervals is then found to be governed by a Floquet Hamiltonian, having exponentially localized eigenstates in the localized phase. Thus, it came to be accepted that the localization-delocalization transition in more than one dimensional Anderson problem can manifest in the rotor dynamics if the `temporal' dimension is increased, i.e., when driven with multiple incommensurate frequencies. \\

In this work, we show for the first time that a quasi-localization to delocalization transition can also manifest in the QKR driven with a single frequency but where the kick amplitudes at subsequent stroboscopic instants follow a binary Fibonacci sequence. We dub this variant of the QKR  as the \textit{Fibonacci quantum kicked rotor} (FQKR). As our main result, we show the emergence of a `pre-ergodic' regime in the limit of high drive frequencies during which the wave-function of the FQKR remains `dynamically quasi-localized' in the angular momentum space. Although the dynamics eventually becomes ergodic, the pre-ergodic regime has a long experimentally relevant lifetime, which further increases with the kicking frequency.  However, at lower drive frequencies, the dynamics is always found to be ergodic. We note that the Fibonacci drive  \cite{morse1921, morse1921_b} has been extensively used to  explore the consequences of temporal quasi-periodicity in various out-of-equilibrium systems \cite{kadanoff83,ostlund83,sutherland86, nandy17, andrew18, nandy18, maity19, sayak19, mukherjee20, pengfei20, zhao21}. \\


\begin{figure*}
	\subfigure[]{
		\includegraphics[width=0.45\linewidth,height=0.35\linewidth]{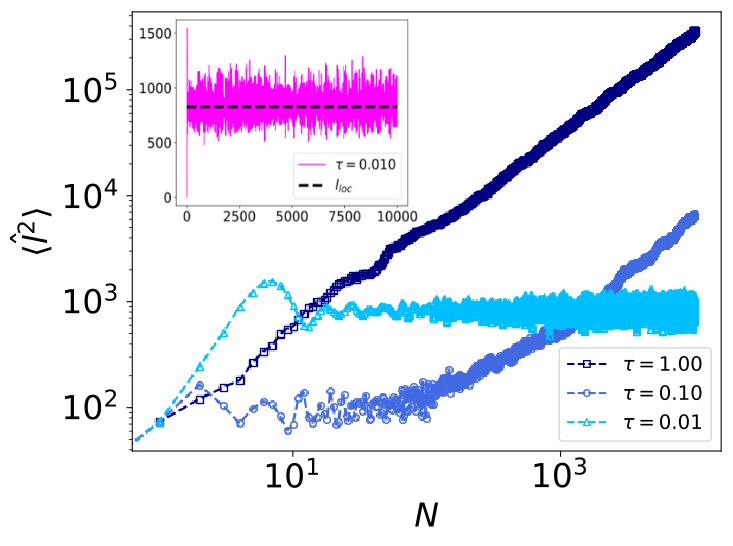}\label{fig_fibo_freq}}
	\subfigure[]{
		\includegraphics[width=0.45\linewidth,height=0.35\linewidth]{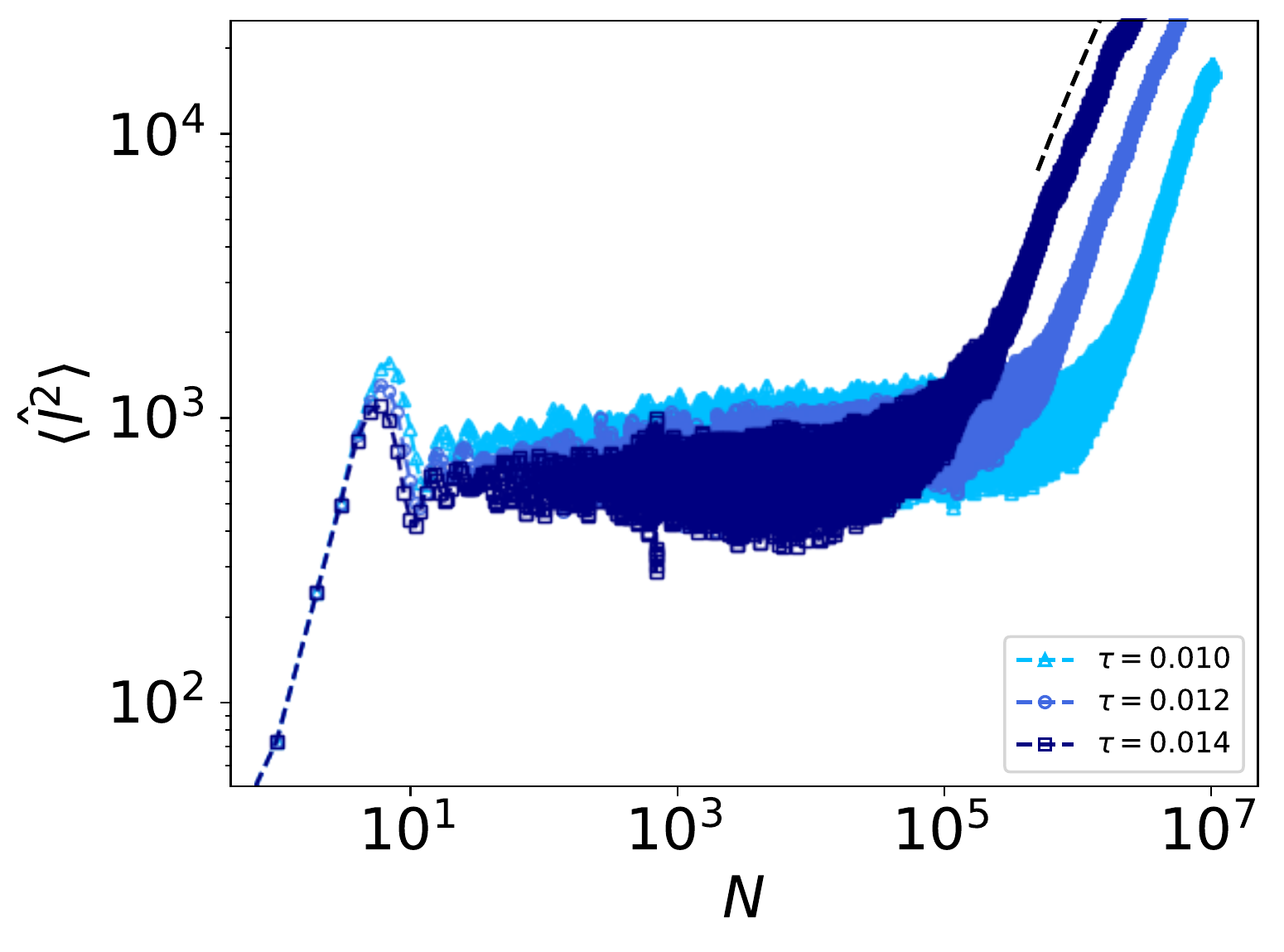}\label{fig_fibo_long}}
	\caption{{(a) The kinetic energy of a FQKR grows diffusively at lower drive frequencies and dynamically quasi-localizes at high frequency ($\tau<0.1$). The inset shows that the fluctuations in the quasi-localized (pre-ergodic) regime are evenly spread about the mean kinetic energy calculated from the perturbation analysis (black dashed line ) using Eq.~\eqref{eq_approx_loc}. (b) The quasi-localization is destroyed and an ergodic behavior is found to emerge at very long times. This  can be seen from the linear growth of the kinetic energy with unit slope for $N>10^5$. The (black) dashed line with unit slope is provided for visual reference. Further, the time after which localization is destroyed progressively increases as the frequency ($\tau^{-1}$) is increased .  The kick amplitudes chosen for the plot are $\mathcal{K}_1=10$, $\mathcal{K}_2=12$ and the initial state is chosen to be the angular momentum eigenstate $\ket{\psi_0}=\ket{l=0}$.}}
\end{figure*}
An important distinction in the approach of our work from previously studied rotors driven at incommensurate frequencies  \cite{casati89, ringot00},  is that the quasi-periodicity in our model is externally enforced through a binary sequence of kick amplitudes. This is manifestly different from previous works where, for example, quasi-peroidicity is indirectly enforced through quasi-periodic phase shifts of the position operator, keeping the kick amplitude  constant \cite{casati89}. Secondly, the quasi-localization observed in the pre-ergodic regime is not discernible from a standard Floquet analysis of the evolution operator. Rather, it follows from the existence of \textit{self-similar eigenstates} and is thus different from hitherto observed dynamical localization in the conventional QKR.
Crucially, we note that the existence of a pre-ergodic regime is unique to the FQKR that, to the best of our knowledge, has not been reported elsewhere. { Further, we have verified that the quasi-localization is  not found for aperiodic  binary sequences.  (see Appendix~\ref{app_birand} for details).}\\
 
The rest of the article is organized as follows: The FQKR model is introduced in Sec.~\ref{sec_model}. In Sec.~\ref{sec_num}, we present results from the numerical simulation of the dynamics of the FQKR. The quasi-localization behavior observed is explained through a perturbative analysis of the time-evolution operator in Sec.~\ref{sec_perturb}. Finally, we summarize our results in Sec.~\ref{sec_summary}. In addition, five appendices are provided a the end which recapitulates known results, especially that of the regular QKR, and also outline the detailed calculations needed to arrive at some of the results presented in this article.

\section{Model}\label{sec_model}
The QKR is represented by the Hamiltonian, 
\begin{equation}\label{eq_hamil}
H(t)=\frac{\hat{l}^2}{2I}+\cos{\hat{\theta}}\sum_{N=0}^\infty K_N\delta\left(t-NT\right),
\end{equation}
where $\hat{\theta}$ and $\hat{l}$ are the angular displacement and angular momentum operators, respectively, while $I$ is the moment of inertia of the rotor. 
The rotor evolves {freely with time period $T$} between subsequent kicks {of strength $K_N$}, which act on the rotor at the stroboscopic instants $NT$, where {$N\in\mathbb{Z^+}$}. 
However, in our case, we consider  $K_N\in\{\mathcal{K}_1,\mathcal{K}_2\}$ and thus $\{K_N\}_{N=1}^\infty \equiv K_1, K_2,\dots$ constitutes a binary sequence.\\

\begin{figure*}
	\subfigure[]{
		\includegraphics[width=0.45\linewidth]{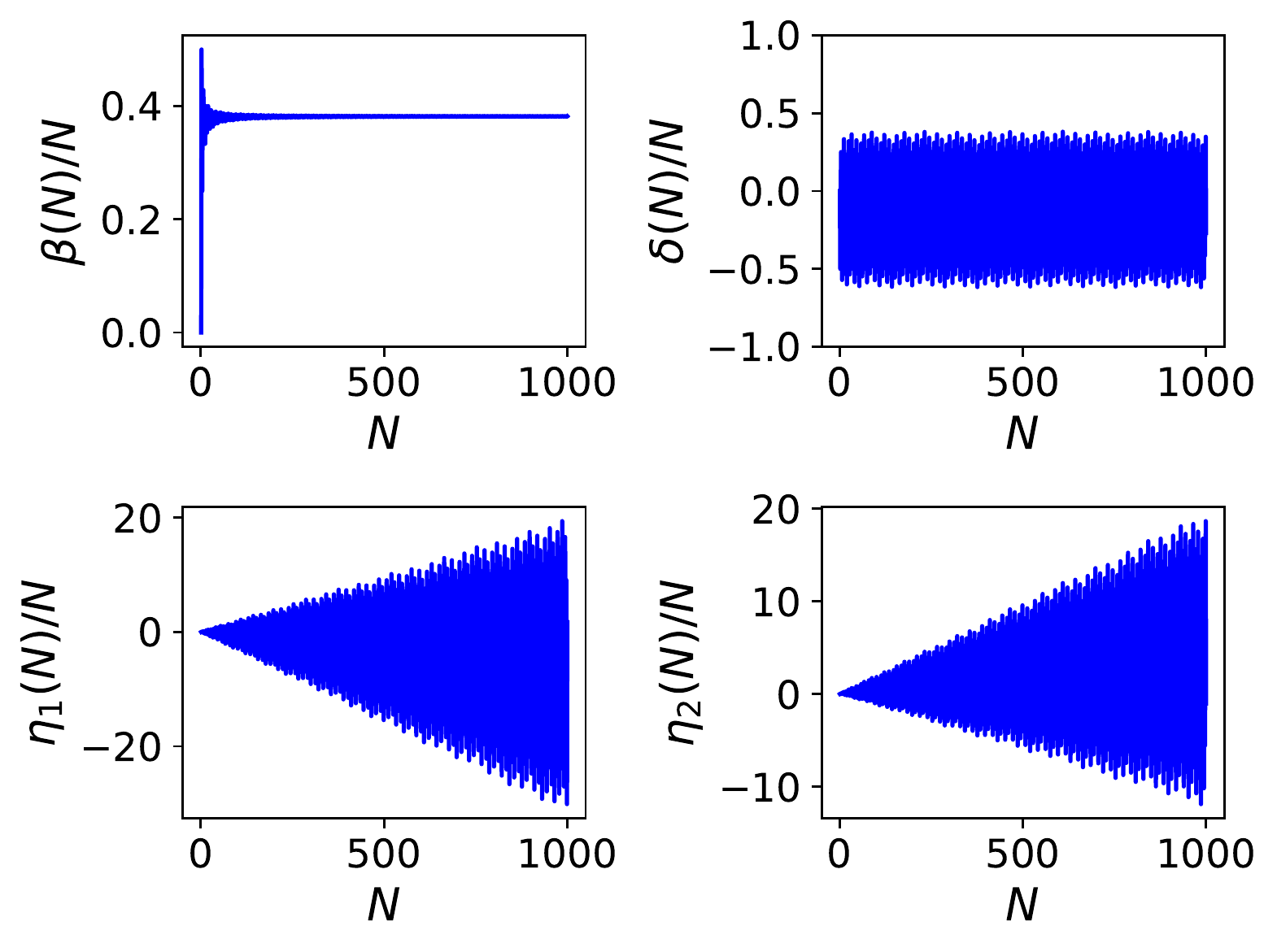}\label{fig_coeff_strob}}	
	\hspace{1cm}
	\subfigure[]{
		\includegraphics[width=0.45\linewidth]{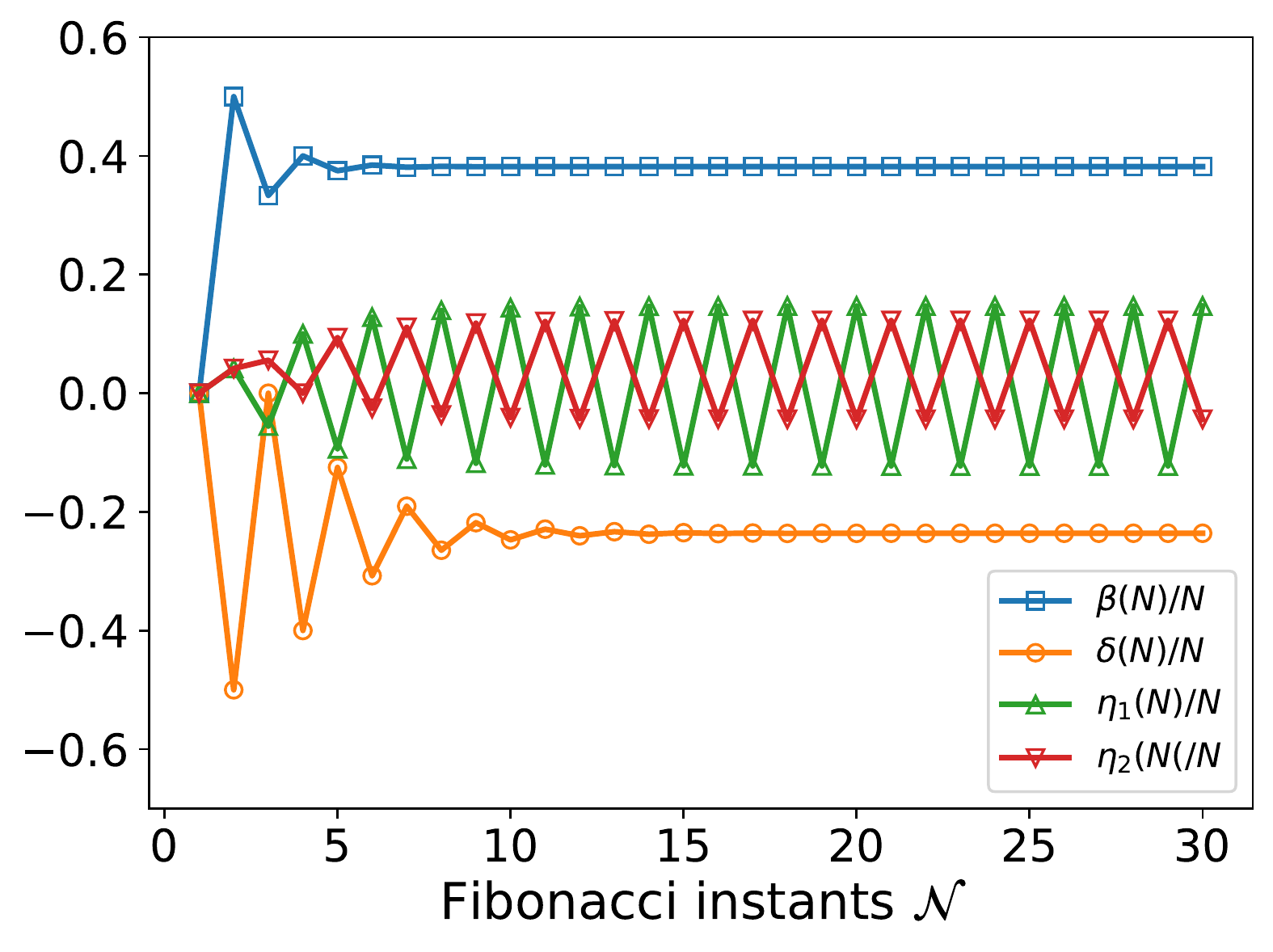}\label{fig_coeff_fibo}}	
	\caption{The normalized expansion coefficients defined in Eq.~\eqref{eq_eff_hamil} at (a) stroboscopic instants $N$ and (b) Fibonacci instants $\mathcal{N}$.}
\end{figure*}
If the rotor is initially in a state $\ket{\psi(t=0)}=\ket{\psi_0}$, the state of the system at the $N$-th stroboscopic instant  (just before the $N^{th}$ kick) is thus given by,
\begin{equation}\label{eq_psin}
\ket{\psi_N}\equiv\ket{\psi(NT)}=U(NT,0)\ket{\psi_0}=U_NU_{N-1}\dots U_1\ket{\psi_0},
\end{equation}  
{$U_N$ being the the unitary operator propagating the system from the $(N-1)^{th}$ to $N^{th}$ time interval given by},
\begin{equation}\label{eq_uni_single}
U_N = \mathcal{T}\exp\left(-i\int_{(N-1)T}^{NT}H(t)dt\right) = e^{-i\frac{\hat{l}^2}{2}\tau}e^{-iK_N\cos\hat{\theta}},
\end{equation} 
where $\mathcal{T}$ is the time-ordering operator. Note that we have set $\hbar=1$  and rescaled the time-period $T$ to a dimensionless parameter $\tau =  T/I$ \cite{hs_equivalence}.
In the conventional QKR,  $K_N = \mathcal{K}$ for all $N$; the dynamics is then equivalent to an evolution under the Floquet propagator $U_F =e^{-i\frac{\hat{l}^2}{2}\tau}e^{-i\mathcal{K}\cos\hat{\theta}}=e^{-iH_F\tau}$, such that  Eq.~\eqref{eq_psin} can be written as,
$\ket{\psi_N} = e^{-iH_FN}\ket{\psi_0}.$
For a finite $\mathcal{K}$, the eigenstates of the Floquet Hamiltonian $H_F$ are exponentially localized in the angular momentum space. This leads to a localization of the wave-function in the angular momentum space and the dynamics is consequently non-ergodic (see Appendix~\ref{app_regular} for details).\\

Referring to  Eq.~\eqref{eq_hamil}, let us now introduce the FQKR as a quantum rotor 
where the sequence of kicks driving the rotor follows the Fibonacci sequence,

\begin{equation}\label{eq_fibo_gen}
	K_N=\mathcal{K}_1+\left(2-\gamma(N)\right)\frac{\Delta \mathcal{K}}{2},
\end{equation}
where $\Delta \mathcal{K} = \mathcal{K}_2-\mathcal{K}_1$. The generating function is given by  $\gamma(N)=\lfloor (N+1)G\rfloor - \lfloor NG\rfloor$, where $\lfloor x\rfloor$ denotes the greatest integer less than or equal to $x$ and $G$ is the golden mean $G = (\sqrt{5}+1)/2$. $\gamma(N$) can thus assume the values $0$ or $2$. To see that Eq.~\eqref{eq_fibo_gen} generates a Fibonacci sequence, let us look at the sequences generated at the \textit{Fibonacci instants} $\mathcal{N}\in\mathcal{Z}^+$. The stroboscopic instant corresponding the $\mathcal{N}^{th}$ Fibonacci instant is given by the function $F(\mathcal{N})$ which satisfies $F(\mathcal{N}) = F(\mathcal{N}-1)+ F(\mathcal{N}-2)$. Thus, we have $F(1) = 1$, $F(2) = 2$, $F(3) = 3$, $F(4) = 5$, $F(5) = 8$, $F(6) = 13$, and so on.
Assuming that $K_1 = \mathcal{K}_1$, we therefore find,
\begin{flalign}\label{eq_fibo_seq}
\{K_N\}_{N=1}^{F(1)} \equiv S_1=&\mathcal{K}_1,\non\\
\{K_N\}_{N=1}^{F(2)} \equiv S_2=&\mathcal{K}_1~\mathcal{K}_2,\non\\
\{K_N\}_{N=1}^{F(3)} \equiv S_3=&\mathcal{K}_1~\mathcal{K}_2~\mathcal{K}_1,\non\\
\{K_N\}_{N=1}^{F(4)} \equiv S_4=&\mathcal{K}_1~\mathcal{K}_2~\mathcal{K}_1~\mathcal{K}_1~\mathcal{K}_2,\non\\
\{K_N\}_{N=1}^{F(5)} \equiv S_5=&\mathcal{K}_1~\mathcal{K}_2~\mathcal{K}_1~\mathcal{K}_1~\mathcal{K}_2~\mathcal{K}_1~\mathcal{K}_2~\mathcal{K}_1,\non\\
\end{flalign}
where $S_n$ denotes the $n^{th}$ Fibonacci sequence, satisfying $S_n=S_{n-1}S_{n-2}$.

\section{Numerical results}\label{sec_num}
For numerical simulations,  the initial state is chosen to be a normalized Gaussian quantum state centered around the angular momentum $l_0$,
	$\psi_0(l) = \left(\frac{2}{\pi}\right)^{\frac{1}{4}}e^{-(l-l_0)^2}$,
unless mentioned otherwise. Further, we employ a truncated basis of angular momentum states for the numerics so that $(l_0-R/2)\leq l\leq (l_0+R/2-1)$, where $R$ is chosen to be large enough to ensure normalization of the wave function at all times. For dynamical localized wave-functions, this is ensured by choosing $R\gg\xi$, where $\xi$ is the localization length. The temporal evolution of the kinetic energy $\langle l^2\rangle$ for the  FQKR with $l_0=0$ is shown in Fig.~\ref{fig_fibo_freq}. We find that the kinetic energy grows diffusively at low frequencies (high $\tau$) and localizes at high, but finite frequency $\tau\leq 0.01$. However, as shown in Fig.~\ref{fig_fibo_long}, the localization does not persist indefinitely and a diffusive behavior emerges eventually. Nevertheless, the time after which this happens rapidly increases as the drive frequency increases. We have numerically verified that the results remain qualitatively the same for different choices of $\mathcal{K}_1$ and $\mathcal{K}_2$. Further, as shown in Appendix~\ref{app_diff_l0}, the above results remain qualitatively the same for different choices of $l_0$.



\section{Perturbative analysis}\label{sec_perturb}
The quasi-localized behavior observed in the limit of high frequencies suggests that there might possibly exist an effective Hamiltonian, similar to the Floquet Hamiltonian, which governs the dynamics of the FQKR in this limit. Importantly, the eigenstates of this effective Hamiltonian should also be exponentially localized  in the angular momentum space. To verify if this is indeed the case,  we resort to a perturbative analysis of the unitary evolution operator, with $\tau$ as the small parameter.

\subsection{Perturbative expansion of the unitary operator}
For $\tau\ll 1$, the unitary operator $U_N$ driving the evolution between the $(N-1)^{th}$ and $N^{th}$ kicks can be written as, 
$U_N = e^{-i\frac{\hat{l}^2}{2}\tau}e^{-iK_N\cos\hat{\theta}} \approx e^{-iL_{1,2}}$,
where $L_{1,2}$ is calculated from a Baker-Campbell-Hausdorff (BCH) expansion of $U_N$,
\begin{multline}\label{eq_L}
L_{1,2}=\mathcal{K}_{1,2}\cos\hat{\theta}+\frac{\tau}{2}\Big[\hat{l}^2+\frac{\mathcal{K}_{1,2}}{2}\Big(\hat{l}\sin(\hat{\theta})+\sin(\hat{\theta})\hat{l}\Big)\\+\frac{\mathcal{K}_{1,2}^2}{12}\sin^2(\hat{\theta})\Big]+\mathcal{O}(\tau^2), 
\end{multline}
having retained terms only up to linear order in $\tau$.
It can be shown (see Appendix~\ref{app_bch}) that the propagator driving the evolution up to the  $N^{th}$ stroboscopic instant  assumes the form $U(N,0)=e^{-iH_{N}N}$, where the Hamiltonian $H_N$ takes the form,
\begin{multline}\label{eq_eff_hamil}
H_{N}=\frac{\alpha(N)}{N}L_1+\frac{\beta(N)}{N}L_2+\frac{\delta(N)}{N}[L_2,L_1]\\+\frac{\eta_1(N)}{N}[L_1,[L_1,L_2]]+\frac{\eta_2(N)}{N}[L_2,[L_2,L_1]]+\mathcal{O}(\tau^2),
\end{multline}
The coefficients $\alpha(N)/N$, $\beta(N)/N$, $\delta(N)/N$, $\eta_1(N)/N$ and $\eta_2(N)/N$, henceforth referred to as the \textit{normalized expansion coefficients} (NECs), depend on the exact form of the binary sequence. The NECs for the Fibonacci sequence have been evaluated in detail in Appendix~\ref{app_bch}.\\

\begin{figure}	
\includegraphics[width=\linewidth]{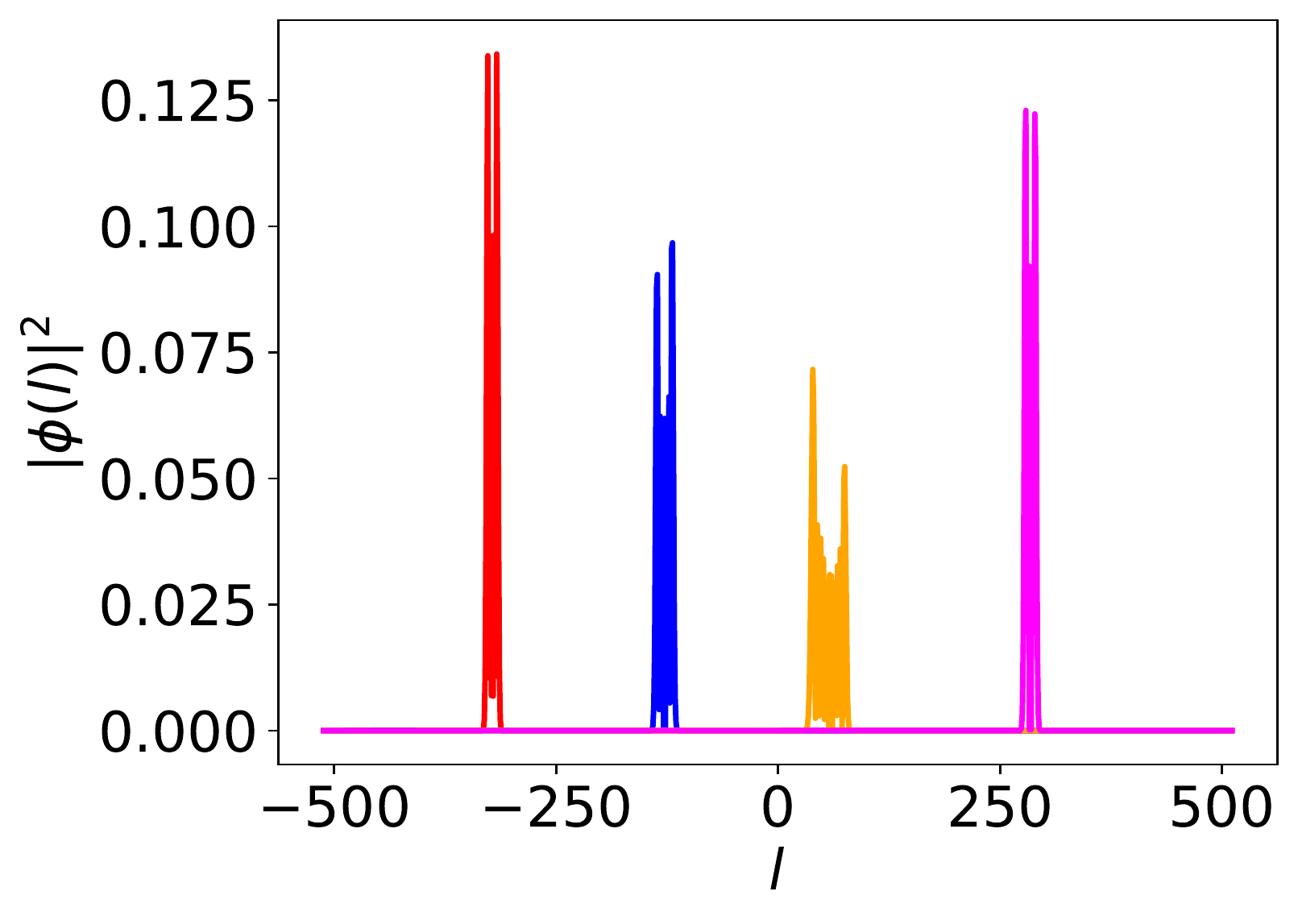}
\caption{Typical eigenstates of the effective Fibonacci Hamiltonian $H_{fi}$, peaked around different values of $l$, for $\mathcal{K}_1 = 10$, $\mathcal{K}_2 = 12$ and $\tau = 0.01$. The eigenstates are found to be localized in the angular momentum space.}\label{fig:eigen} 
\end{figure}

From Fig.~\ref{fig_coeff_strob}, we see that while $\beta(N)/N$ (also $\alpha(N)/N=1-\beta(N)/N$,  as shown in Appendix~\ref{app_bch}) saturates to a constant value, $\delta(N)/N$ exhibits small amplitude fluctuations about a mean value. On the other hand, the growth of the coefficients $\eta_1(N)/N$ and $\eta_2(N)/N$ with $N$ is unbounded. The Hamiltonian $H_{N}$, as defined in Eq.~\eqref{eq_eff_hamil}, is therefore $N$ - dependent and is not a conserved quantity unlike the Floquet Hamiltonian. {However, a different picture emerges if we observe the behavior of the NECs at Fibonacci instants $\mathcal{N}$. In this case, as shown in Fig.~\ref{fig_coeff_fibo}, the coefficients $\beta(F(\mathcal{N}))/F(\mathcal{N})$, $\alpha(F(\mathcal{N}))/F(\mathcal{N})$ and $\delta(F(\mathcal{N}))/F(\mathcal{N})$ are found to saturate to steady values for $\mathcal{N}>10$ while the coefficients $\eta_1(F(\mathcal{N}))/F(\mathcal{N})$ and $\eta_2(F(\mathcal{N}))/F(\mathcal{N})$ oscillate between a pair of constant values. Thus, for $\mathcal{N}>10$, one can substitute $H_{N=F(\mathcal{N})}=H_{fi}$ (see Appendix~\ref{app_eff} for details), where $H_{fi}$ is an \textit{effective Fibonacci Hamiltonian} which governs the dynamics of the rotor at Fibonacci instants. Further, as shown in Fig.~\ref{fig:eigen}, the eigenstates of $H_{fi}$ are also localized in the angular momentum basis. It therefore follows that the dynamics of FQKR should mimic that of the regular QKR when observed at Fibonacci instants $N=F(\mathcal{N})$ and when terms of order $\mathcal{O}(\tau^2)$ can be neglected in the perturbative analysis. Nevertheless, the existence of an effective Fibonacci Hamiltonian at Fibonacci instants does not guarantee the persistent localization seen throughout the evolution. 
In fact, the localization in between two subsequent Fibonacci instants is particularly surprising, given that the growth of the coefficients $\eta_1(N)/N$ and $\eta_2(N)/N$ are unbounded when observed at stroboscopic instants, as already seen in Fig.~\ref{fig_coeff_strob}.\\

This apparent contradiction can be explained if one inspects the self-similar or fractal nature of the Fibonacci sequence. To elaborate, let us consider the evolution between two subsequent Fibonacci instants $\mathcal{N}^*$ and $\mathcal{N}^*+1$, where we assume $F(\mathcal{N}^*)\gg1$ so as to ensure that the NECs have saturated to their mean values. 
From Eq.~\eqref{eq_fibo_seq}, we note that the sequence of kicks up to $N^* = F(\mathcal{N}^*)$ and $N^{**} = F(\mathcal{N}^*+1)$ is given by $S_{\mathcal{N}^*}$ and $S_{\mathcal{N}^*+1}$, respectively. However, by construction, $S_{\mathcal{N}^*+1} = S_{\mathcal{N}^*}S_{\mathcal{N}^*-1}$, which immediately implies that the sequence of kicks between $N^*$ and $N^{**}$ is nothing but the sequence $S_{\mathcal{N}^*-1}$.\\
	
The dynamics of the rotor in between two Fibonacci instants can now be analysed as follows. Given a localized wave-function $\ket{\psi_{N^*}}$, the evolution of this wave-function in between the Fibonacci instants $\mathcal{N}^*$ and $\mathcal{N^*}+1$ is generated by the sequence of kicks $S_{\mathcal{N}^*-1}$. Let us now denote the Fibonacci instants nested within the sequence $S_{\mathcal{N}^*-1}$ as $\mathcal{M}$, where $\mathcal{M}=0,1,2,\dots,\mathcal{N}^*-1$. The wave-function of the rotor at these instants thus satisfies, $\ket{\psi_{N^*+F(\mathcal{M})}} \approx U_{fi}^{F(\mathcal{M})}\ket{\psi_{N^*}}$.
Hence, it straightaway follows that the wave-function also remains localized at the instants $N^{*}+F(\mathcal{M})$, as the evolution is driven by the same effective Fibonacci Hamiltonian $U_{fi}$. Proceeding similarly, one can argue that the sequence of kicks acting between $N = N^*+F(\mathcal{M}^*)$ and $N^{*}+F(\mathcal{M}^*+1)$ is precisely the sequence $S_{\mathcal{M}^*-1}$ and hence the wave-function remains localized at the Fibonacci instants nested between $N = N^*+F(\mathcal{M}^*)$ and $N^{*}+F(\mathcal{M}^*+1)$. Thus, the quasi-localization is enforced in a self-similar way between two subsequent Fibonacci instants. 
In other words, the localised eigenstates of $U_{fi}$  act as \textit{self-similar eigenstates} of $U(N,0)$ and ultimately lead to the quasi-localized dynamics observed stroboscopically. \\

{To further support the arguments presented above, we estimate the localization length $\langle\hat{l}^2\rangle_{loc}$, with the initial state of the rotor being an angular momentum eigenstate, $\ket{\psi_0}=\ket{l_0}$, for simplicity in calculation. Assuming that, $\ket{\psi_{F(\mathcal{N})}}\approx U_{fi}^{F(\mathcal{N})}\ket{l_0}$ and $U_{fi}=VDV^{\dagger}$, where $D$ is a diagonal matrix, it can be shown that (see Appendix~\ref{app_regular}),
\begin{equation}\label{eq_approx_loc}
	\langle\hat{l}^2\rangle_{loc} = \sum_{l,m}l^2\left|V_{l_0m}\right|^2\left|V_{lm}\right|^2
\end{equation} 
The localization length calculated using the above equation  indeed turns out  to be similar to that obtained from exact numerics. This is illustrated in the inset of Fig.~\ref{fig_fibo_long} where the fluctuations in the kinetic energy for $N<10^4$ is found} to be evenly distributed about the mean value calculated using Eq.~\eqref{eq_approx_loc}.\\ 

It is important to realize that higher order terms in the BCH expansion, which we have ignored so far, can become significant under two conditions -- (i) if the drive frequency is lowered, so that terms of order $\mathcal{O}(\tau^2)$ becomes significant and (ii) if the NECs of the higher order commutator terms grow boundlessly so that such terms, although initially insignificant, start to dominate after a certain time has elapsed. As we shall see below, the second condition is particularly important as it explains both the ergodic behavior observed at low frequencies and the breakdown of the localization after sufficiently long time at high frequencies.\\
\subsection{Emergence of diffusive behavior at low frequencies}\label{subsec_low_freq}
%
Let us recall that the saturation of the NECs to steady values for $\mathcal{N}\gg1$ is crucial for the existence of an effective Fibonacci Hamiltonian. Without explicitly determining all the commutator terms that may contribute when terms of order $\sim\mathcal{O}(\tau^2)$ are included, let us analyze the expansion coefficients of the commutators $[L_1,[L_1,[L_1,L_2]]]$, $[L_2,[L_2,[L_1,L_2]]]$ and $[L_1,[L_2,[L_1,L_2]]]$. We denote the corresponding expansion coefficients as $\mu_1(N)$, $\mu_2(N)$ and $\mu_3(N)$, respectively. These NECs at $N=F(\mathcal{N})$ with $\mathcal{N}\gg1$ are found to be \cite{andrew18},
\begin{widetext}
\begin{subequations}\label{eq_coeff_fourth_order}
\begin{equation}
	\frac{\mu_1(F(\mathcal{N}))}{F(\mathcal{N})}=\frac{(-1)^\mathcal{N}}{120}\left[G^{\mathcal{N}-1}+\frac{1}{G}\left[(-1)^\mathcal{N}\left(3G-4\right)-1-3G\right]\right],
\end{equation}
\begin{equation}
	\frac{\mu_2(F(\mathcal{N}))}{F(\mathcal{N})}=\frac{(-1)^\mathcal{N}}{120}\left[G^{\mathcal{N}-1}\left(2-G\right)+\frac{1}{G}\left[(-1)^\mathcal{N}\left(4G-7\right)-2-G\right]\right],
\end{equation}
\begin{equation}
	\frac{\mu_3(F(\mathcal{N}))}{F(\mathcal{N})}=\frac{(-1)^\mathcal{N}}{120}\left[2G^{\mathcal{N}-1}\left(1-G\right)+\frac{1}{G}\left[(-1)^\mathcal{N}\left(3-G\right)+3+4G\right]\right],
\end{equation}
\end{subequations}
\end{widetext}
where we have ignored terms of order $G^{-(\mathcal{N}+1)}$, $G^{-(2\mathcal{N}+1)}$, $G^{-(3\mathcal{N}+1)}$, etc.
}
It is immediately clear that the NECs defined above do not saturate to steady values even in the asymptotic limit, rather their growth with $\mathcal{N}$ is unbounded. Indeed, it can be shown that this is true for all the NECs of higher order nested commutators \cite{andrew18}. Thus, we conclude that when the frequency of the drive is low enough such that the contribution of higher order terms become significant, the dynamics at Fibonacci instants is no longer governed by an effective Fibonacci Hamiltonian. The evolution of the rotor then mimics that of random driving, thereby leading to the emergence of ergodic behavior after sufficiently long times.

\subsection{Crosssover from pre-ergodic to ergodic regime at high frequencies}\label{subsec_high_time_stability}
The unbounded growth of the NECs in Eq.~\eqref{eq_coeff_fourth_order} has a more important consequence. For $\tau\ll 1$, the higher order terms containing the commutators corresponding to these NECs are insignificant when $\mathcal{N}$ is not too large. However, it is easy to see that there will exist a long but finite time after which such terms will become significant and consequently, the ergodic behavior will set in. This leads to the breakdown of the localization observed in the limit of high frequency. Indeed, one can perform an order of magnitude estimation of the time $N_{deloc}$, after which the diffusive growth is expected to manifest  as follows. From Eq.~\eqref{eq_coeff_fourth_order},  we note that the leading order term unbounded in $\mathcal{N}$ grows as $G^{\mathcal{N}}/120$. The $\mathcal{O}(\tau^2)$ terms in the expansion of Eq.~\eqref{eq_eff_hamil}
thus become significant when $\tau^2G^{\mathcal{N}_{deloc}}/120\sim 1$. For $\tau\approx 0.01$, this translates to $\mathcal{N}_{deloc}\approx 29$ or $N_{deloc}\approx1.3\times10^6$. Thus, within the experimentally realisable time scale, one should observe the quasi-localisation. This agrees remarkably well with the results found from exact numerical calculations (see Fig.~\ref{fig_fibo_long}).

Finally, we note that the delocalization time $\mathcal{N}_{deloc}$ can be interpreted as the time required by the system to `resolve the randomness' of the quasi-periodic drive. The existence of the  time-independent effective Hamiltonian for small $\tau$, combined with the self-similarity of the drive, leads to the an effective periodic evolution from the system's point of view. Only at later times, when higher order terms start to dominate and the effective time-independent Hamiltonian picture breaks down, does the system realise that the drive is not periodic and diffusive dynamics sets in.

\section{Summary}\label{sec_summary}
In conclusion, we have demonstrated that a quantum rotor driven with a binary Fibonacci sequence can exhibit both diffusive and quasi-localized behavior. The latter manifests in the limit of high frequency of the drive, although diffusive behavior eventually sets in after sufficiently long times. It is interesting to note that the pre-ergodic regime, within which the quasi-localization persists at high frequencies, is reminiscent of the \textit{prethermal} regimes observed in out-of-equilibrium many-body quantum systems. In such systems, the presence of approximately conserved quantities prevent the system from thermalizing for a long period of time. An important question that arises is that whether the dynamics of the FQKR can be mapped to a real space lattice model describing spatial localization,  just as the dynamics of the regular QKR is mappable to the one-dimensional Anderson problem. If a mapping does exist, it would be interesting to see how the quasi-localization observed in the FQKR manifests in the dual model. This is however beyond the scope of the present work.

\begin{acknowledgments}
We acknowledge Markus Heyl for discussions and for his very useful comments and suggestions. We acknowledge Utso Bhattacharya, Somnath Maity and Anatoli Polkovnikov for useful discussions and comments. Sourav Bhattacharjee acknowledges CSIR, India for financial support. Souvik Bandyopadhyay acknowledges financial support from PMRF, MHRD, India. AD acknowledges financial support from a SPARC program, MHRD, India and SERB, DST, New Delhi, India..
\end{acknowledgments}

\appendix

\section{Regular quantum kicked rotor}\label{app_regular}
\begin{figure*}
	\subfigure[]{
		\includegraphics[width=0.3\linewidth]{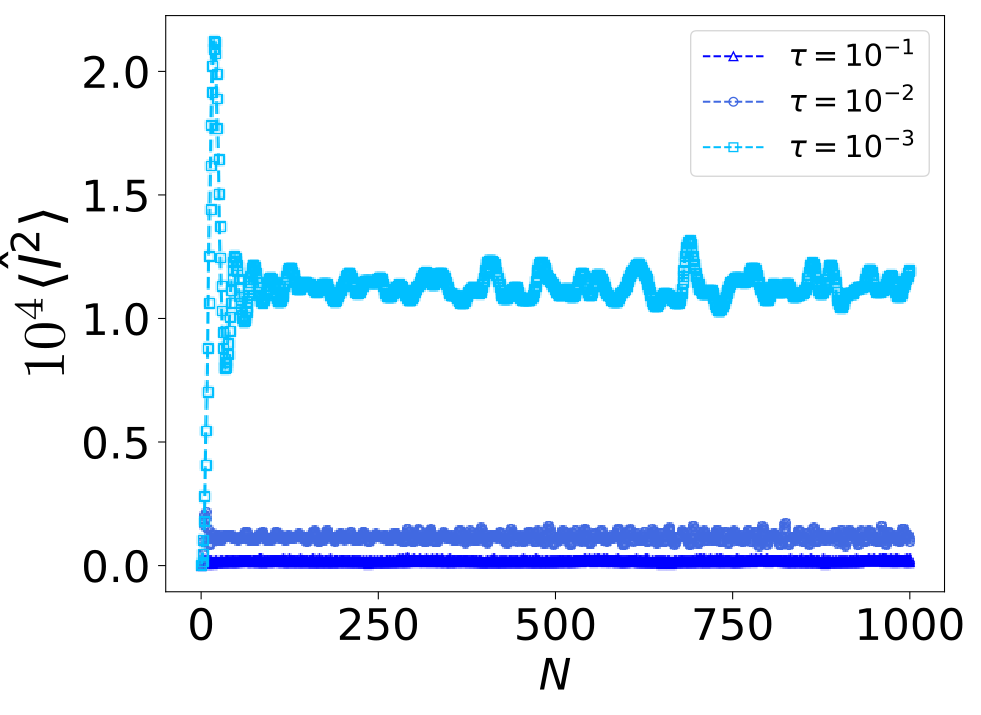}\label{fig_regu}}	
	\subfigure[]{
		\includegraphics[width=0.3\linewidth]{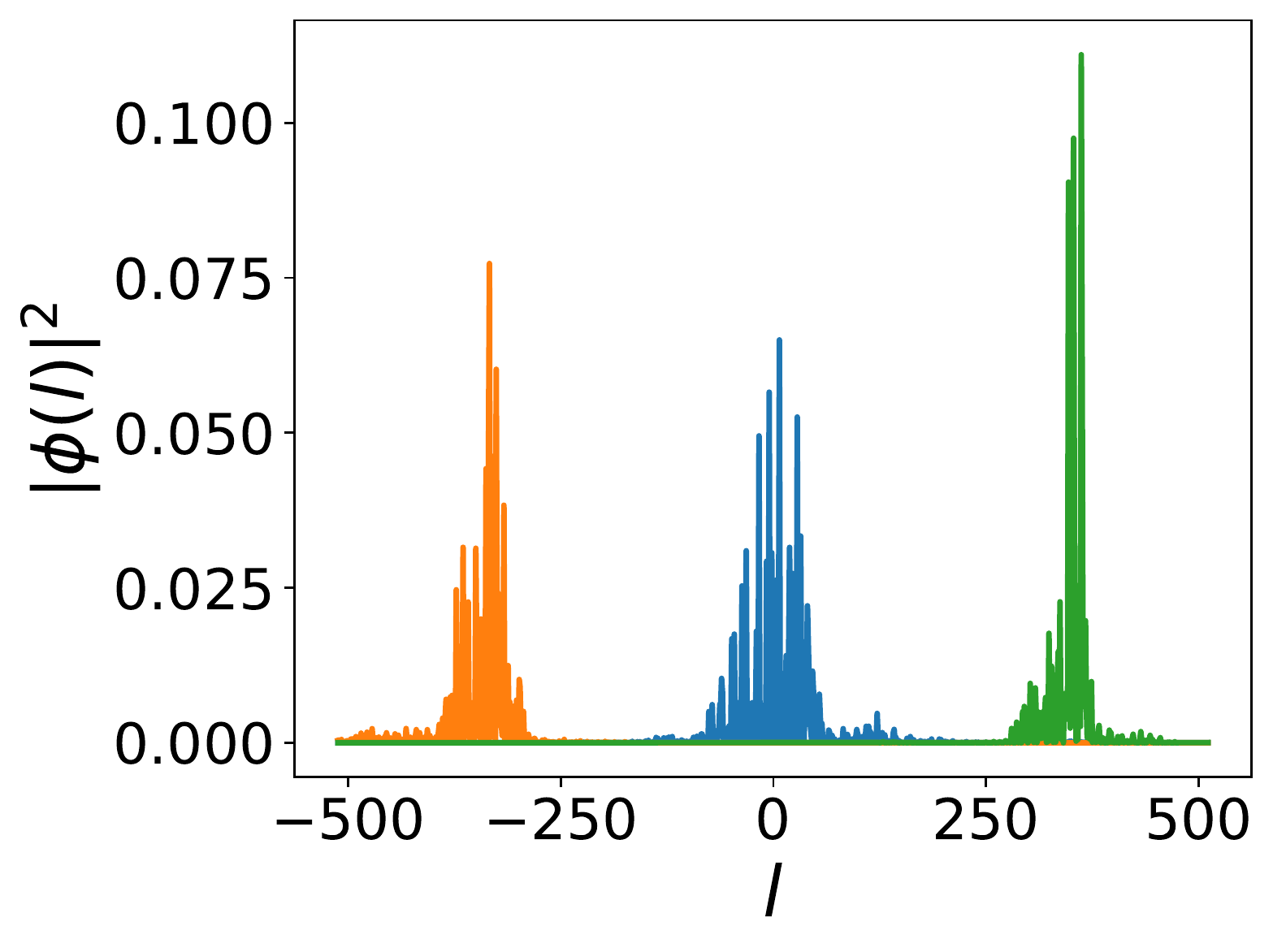}\label{fig_eigen_regu}}
	\subfigure[]{
		\includegraphics[width=0.3\linewidth]{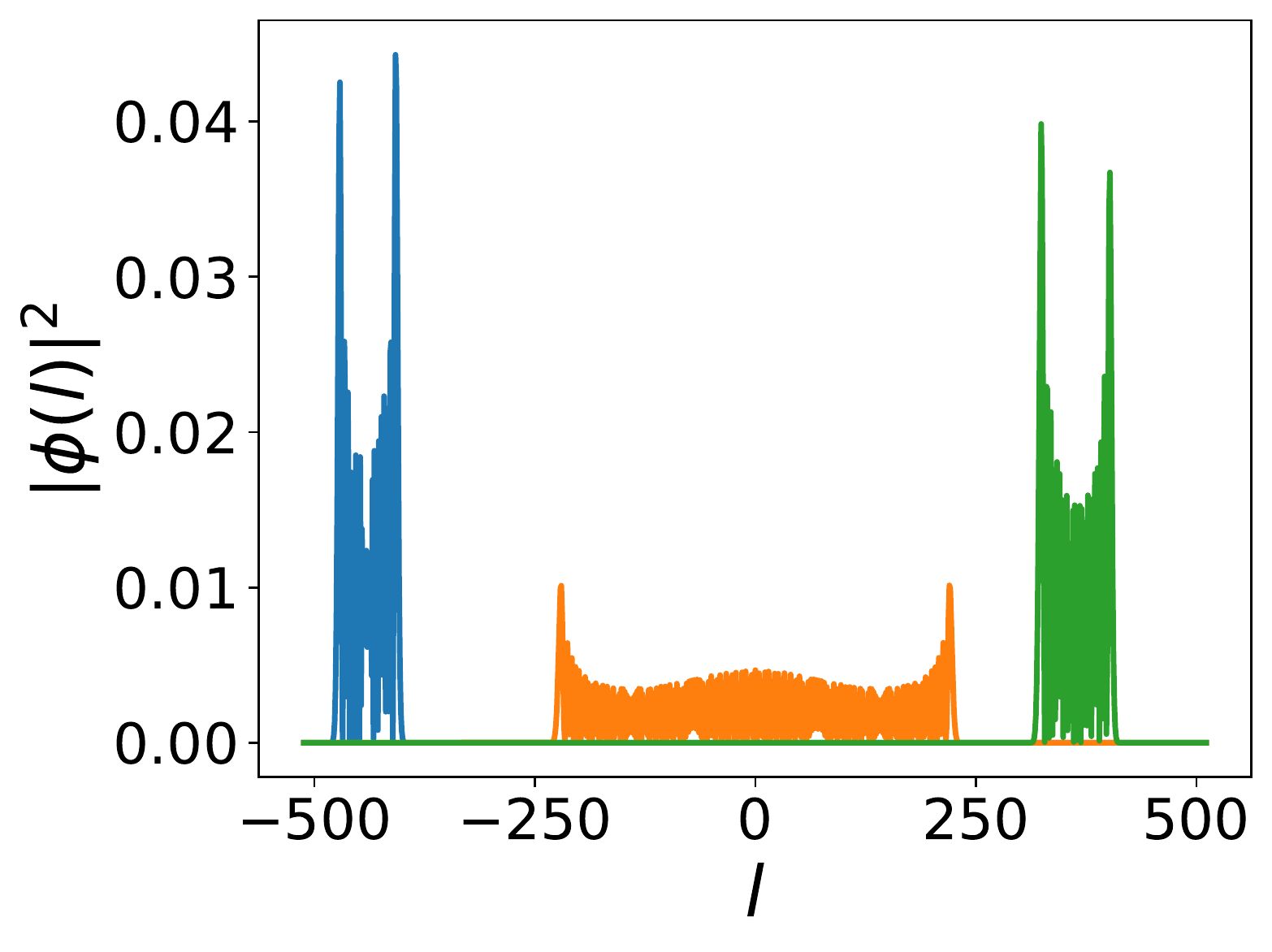}\label{fig_eigen_regu_low}}	
	\caption{(a) Evolution of the kinetic energy observed at stroboscopic instants $N$ for a regular rotor, $\mathcal{K}_2=15$. The energy dynamically localizes for all driving frequencies, barring resonance conditions which are not considered in this work. Typical eigenstates of the Floquet propagator $U_F$ localized around different values of $l$, defined in Eq.~\eqref{eq_app_un}, with $\mathcal{K} = 15$ and kick frequency (b) $\tau = 1$ and (c) $\tau = 0.001$}
\end{figure*}

The regular QKR is represented by the Hamiltonian,
\begin{equation}\label{eq_hamil_reg}
	H(t)=\frac{\hat{l}^2}{2I}+\mathcal{K}\cos{\hat{\theta}}\sum_{N=0}^\infty \delta\left(t-NT\right).
\end{equation}
As discussed in Sec.~\ref{sec_model}, the above Hamiltonian always exhibits dynamical localization, irrespective of the drive frequency (see Fig.~\ref{fig_regu}). The Floquet propagator governing the evolution of the rotor at stroboscopic instants is given by,
\begin{equation}\label{eq_app_un}
	U_F = e^{-i\frac{\hat{l}^2}{2}\tau}e^{-i\mathcal{K}\cos\hat{\theta}},
\end{equation}
where $\tau=T/I$ is a dimensionless parameter and we have set $\hbar =1$. Let us consider the eigen-spectrum of the Floquet propagator:  $U_F=\sum_m e^{i\phi_m}\ket{\phi_m}\bra{\phi_m}$. As the Hilbert space dimension is infinite snd the \textit{quasi-energies} $\phi_m$ are defined modulo $2\pi$, the Floquet propagator has a dense eigen-spectrum with ill-defined mean level spacing. However, as shown in Fig.~\ref{fig_eigen_regu}, all the eigenstates $\ket{\phi_m}$ turn out to be exponentially localized in the angular momentum space, unless $\tau\ll1$  which we shall consider shortly. To see how these properties lead to a dynamical localization in the dynamics, let us explicitly calculate the kinetic energy in terms of the matrix elements of the Floquet propagator. Without loss of generality, we assume that the rotor is initialized in a definite angular momentum state $\ket{l_0}$. The kinetic energy after $N$ stroboscopic instants can then be evaluated as,  
\begin{widetext} 
\begin{equation}\label{eq_kin_exp}
	\nonumber\langle \hat{l}^2\rangle=\bra{l_0}U_F^{N\dagger}\hat{l}^2U_F^N\ket{l_0}
	=\sum_{l,m,m'}l^2e^{iN\left(\phi_{m}-\phi_{m'}\right)}V_{l_0m'}V_{lm'}^*V_{lm}V_{l_0m}^*, 
\end{equation}
\end{widetext}
where $V_{l_0m} = \braket{l_0|\phi_m}$ and so on. As the eigenstates are exponentially localized in the angular momentum basis, we have $V_{ll'}\approx0$ for $|l-l'|>l_s$, where $l_s$ is the \textit{localization length} and is a measure of the number of Floquet eigenstates which overlaps with each angular momentum state. It is thus clear that in the equation above, only a finite number of eigenstates can contribute to the sum, resulting in the effective quasi-energy spectrum being discreet with a mean level spacing of $2\pi/l_s$. 

The onset of dynamical localization can now be explained as follows. If $2\pi N/l_s\gg1$, all the oscillating terms in Eq.~\eqref{eq_kin_exp} vanish; the average kinetic energy evaluates to,
\begin{equation}
	\langle \hat{l}^2\rangle=\sum_{l,m}l^2\left|V_{l_0m}\right|^2\left|V_{lm}\right|^2\sim l_s^2+l_0^2,
\end{equation}
which is independent of $N$. Further, the Heisenberg time, defined as the initial time for which the kinetic energy grows diffusively, can also be roughly approximated from $2\pi N^*/l_s\approx 1$  or $N^*\sim l_s$. As the kinetic energy is known to follow the classical dynamics till $N^*$ with a diffusion constant $\sim \mathcal{K}_{cl}^2=\mathcal{K}^2\tau^2$, we have 
\begin{equation}
	\langle \hat{l}^2\rangle \sim \mathcal{K}^2\tau^2N^*+l_0^2.
\end{equation}
or,
\begin{equation}
	l_s^2\sim \mathcal{K}^2\tau^2l_s,
\end{equation}
which determines both the localization length and the Heisenberg time as,
\begin{equation}\label{eq_ls}
	l_s\sim \mathcal{K}^2\tau^2.
\end{equation}

\section{Dynamics of the rotor under biperiodic and aperiodic binary sequence}\label{app_birand}
\begin{figure*}
	\subfigure[]{
		\includegraphics[width=0.45\linewidth]{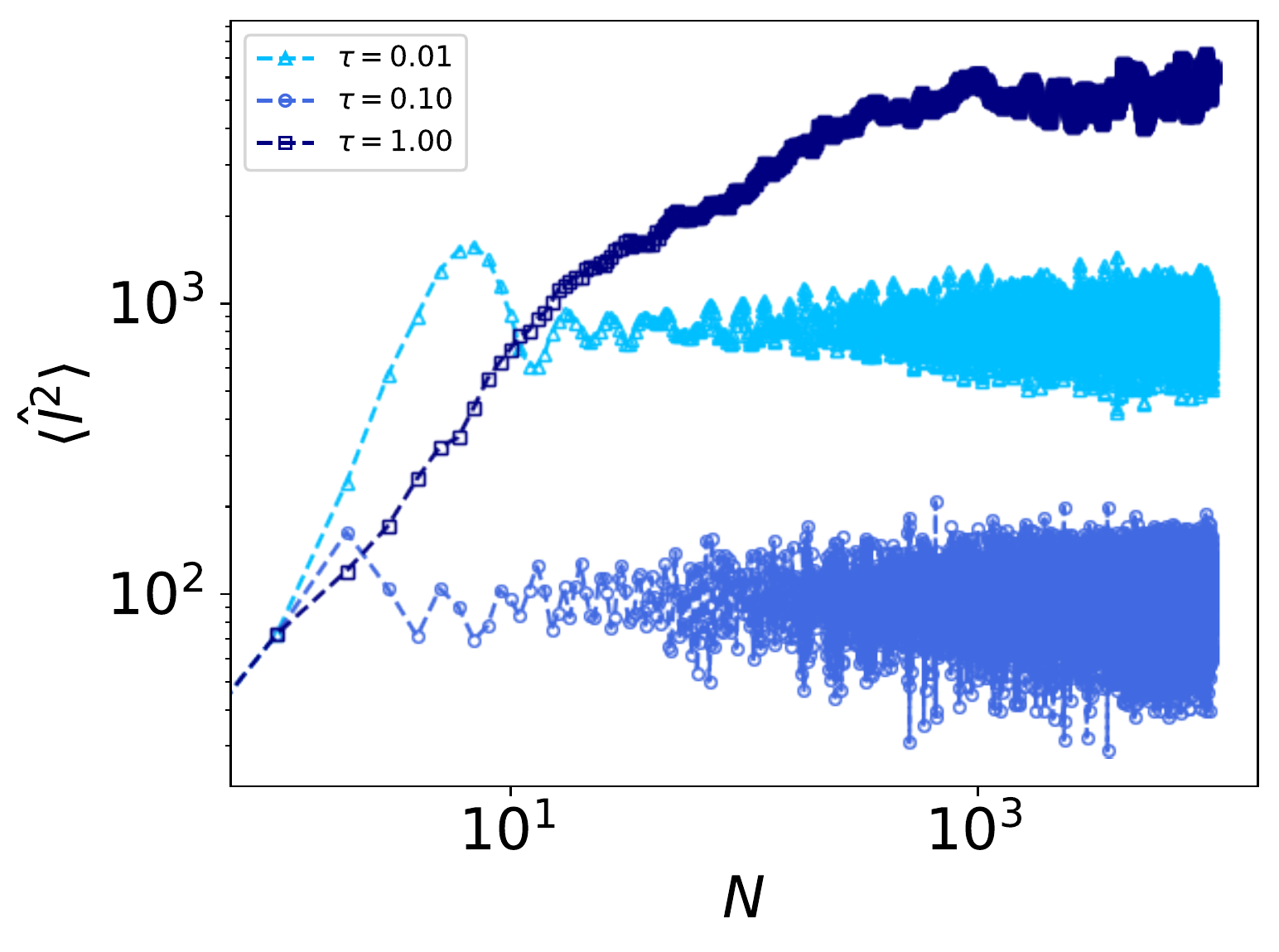}\label{fig_bi}}	
	\subfigure[]{
		\includegraphics[width=0.45\linewidth]{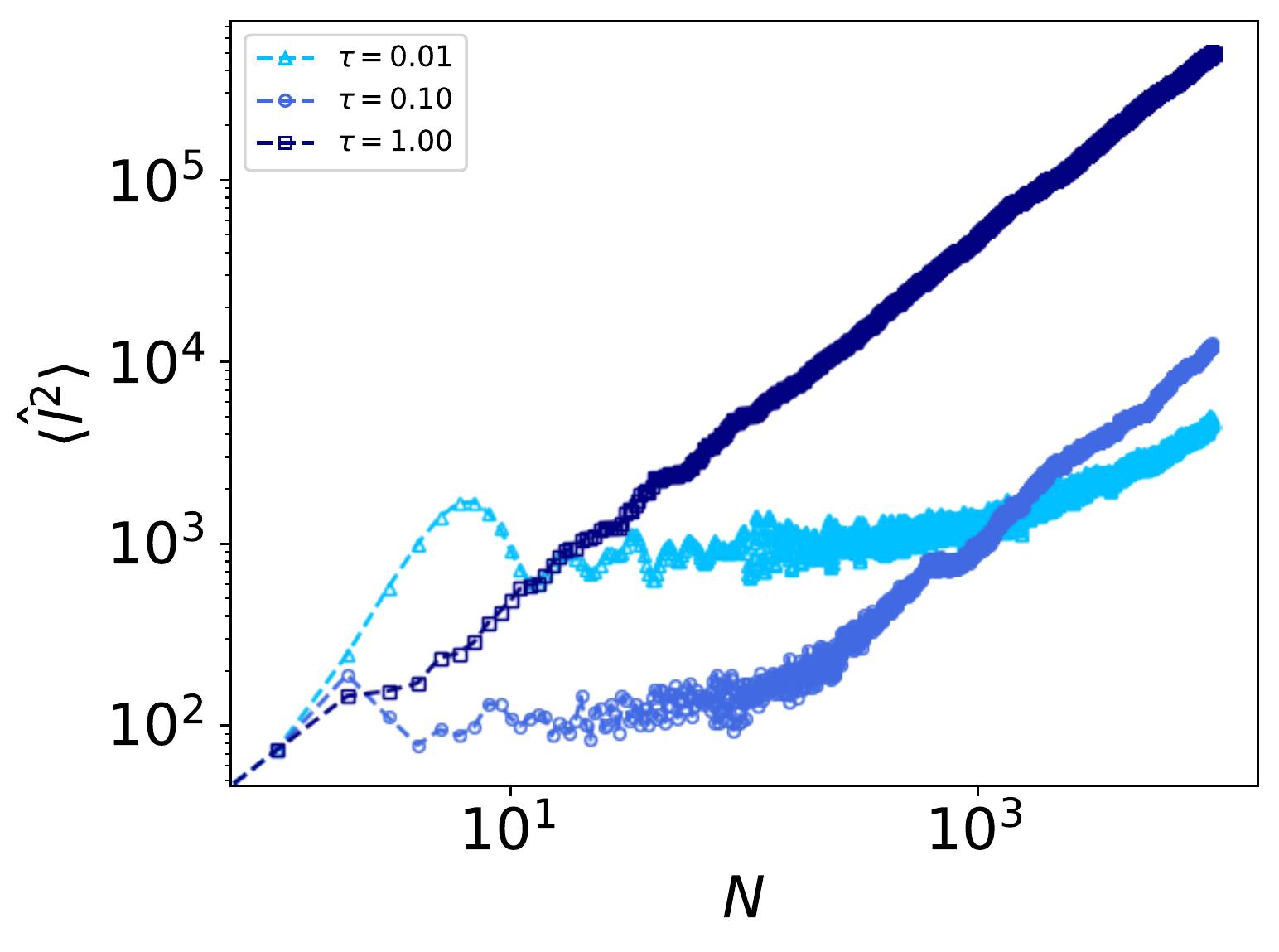}\label{fig_random}}
	\caption{(a) Evolution of the kinetic energy observed at stroboscopic instants $N$ when the rotor is driven with (a) bi-periodic and (b) aperiodic binary sequences and the kick aplitudes chosen as $\mathcal{K}_1=10$ and $\mathcal{K}_2=12$. The energy saturates for all driving frequencies in the case of the bi-periodic sequence while it grows diffusively in the other case.}\label{fig_bi_rand}	
\end{figure*}

The dynamical quasi-localization observed in the case of the FQKR at high frequency drives is not guaranteed for arbitrary binary sequences. {To illustrate this, we consider a QKR driven with a bi-periodic sequence with $K_N=\mathcal{K}_1 (\mathcal{K}_2)$ for even (odd) $N$ and and an aperiodic binary sequence where the amplitude of the kick can either be $\mathcal{K}_1$ or $\mathcal{K}_2$ with equal probability at every stroboscopic instant. This is illustrated in Fig.~\ref{fig_bi_rand} which shows that energy  saturates for all driving frequencies in the case of the bi-periodic sequence (see Fig.~\ref{fig_bi}) while it evolves diffusively in the case of the aperiodic sequence, $\langle \hat{l}^2\rangle \propto N$, irrespective of the driving frequency (see Fig.~\ref{fig_bi})}. 


\section{Quasi-localization dynamics for different $l_0$}\label{app_diff_l0}
\begin{figure}
	\includegraphics[width=0.9\columnwidth]{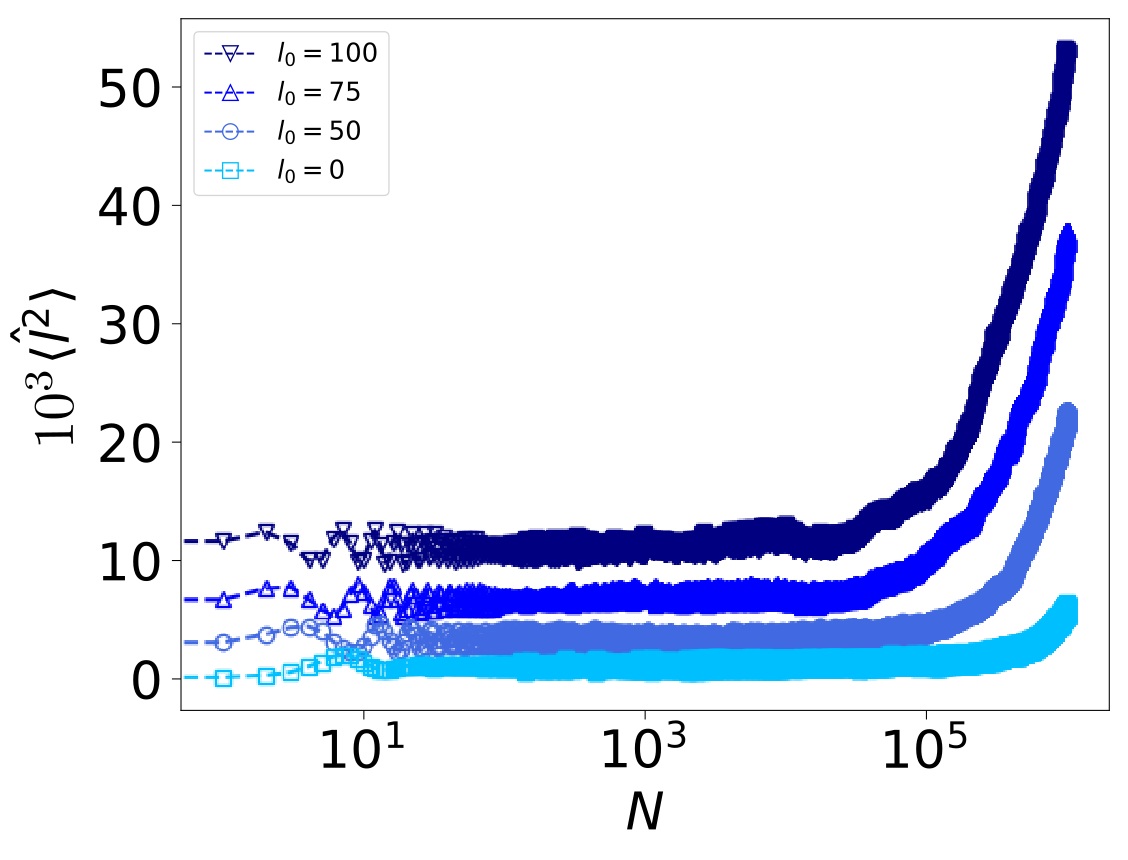}
	\caption{Evolution of the kinetic energy for the FQKR with different choices of $l_0$. The kick amplitudes are chosen to be $\mathcal{K}_1=10$ and $\mathcal{K}_2=12$. }\label{fig_diff_l0}
\end{figure}

As discussed in Sec.~\ref{sec_num}, the initial state of the rotor is chosen to be a coherent Gaussian state centered around the angular momentum $l_0$, i.e., 
$\psi_0(l) = \left(\frac{2}{\pi}\right)^{\frac{1}{4}}e^{-(l-l_0)^2}$. The numerical results presented in the paper correspond to $l_0=0$. To verify that the quasi-localization behavior observed is insensitive to the choice of $l_0$, we plot the evolution of the kinetic energy for different values of $l_0$ in Fig.~\ref{fig_diff_l0}. We find that the choice of $l_0$ only alters the mean value of the kinetic energy in the pre-ergodic phase.

\section{High frequency expansion of the unitary operator}\label{app_bch}

In this section, we shall derive the high frequency expansion of the time-evolution unitary operator for a QKR when driven with a  binary sequence of kicks. As discussed in Eq.~\eqref{eq_uni_single} of Sec.~\ref{sec_model}, the unitary operator driving the evolution between the $(N-1)^{th}$ and $N^{th}$ kicks is given by,
\begin{equation}
	U_n = e^{-i\frac{\hat{l}^2}{2}\tau}e^{-iK_N\cos\hat{\theta}}.
\end{equation} 
We recall that given a pair of non-commuting operators $A$ and $B$, one can write $e^Ae^B=e^C$, where $C$ is given by the Baker-Campbell-Hausdorff formula,
\begin{equation}\label{eq_bch}
	C = A+B+\frac{1}{2}\left[A,B\right]+\frac{1}{12}\Big(\left[A,\left[A,B\right]\right]+\left[B,\left[B,A\right]\right]\Big)+\cdots
\end{equation}
Substituting $A=-i\hat{l}^2\tau/2$ and $B=-iK_N\cos\hat{\theta}$, it is straightforward to check that the only commutators in the above expression which contribute up to linear order in $\tau$ are,
\begin{subequations}
	\begin{equation}
		\left[A,B\right] = -\frac{iK_N\tau}{2}\left(\hat{l}\sin\hat{\theta}+\sin\hat{\theta}~\hat{l}\right)+\mathcal{O}(\tau^2) 
	\end{equation}
	\begin{equation}
		\left[B,\left[B,A\right]\right] = \frac{-iK_N^2\tau}{2}\sin^2\hat{\theta}+\mathcal{O}(\tau^2).
	\end{equation}
\end{subequations}
At high frequencies or $\tau\ll 1$, we can therefore neglect all other commutators in Eq.~\eqref{eq_bch}. Recalling $K_N\in\{\mathcal{K}_1, \mathcal{K}_2\}$ for a binary sequence, we obtain, 
\begin{subequations}\label{eq_fib_approx}
	\begin{equation}
		U_N \approx e^{-iL_{1,2}},
	\end{equation} 
	\begin{multline}
		L_{1,2}=\mathcal{K}_{1,2}\cos\hat{\theta}+\frac{\tau}{2}\Big[\hat{l}^2+\frac{\mathcal{K}_{1,2}}{2}\Big(\hat{l}\sin\hat{\theta}+\sin\hat{\theta}~\hat{l}\Big)\\+\frac{\mathcal{K}_{1,2}^2}{6}\sin^2\hat{\theta}\Big]+\mathcal{O}(\tau^2). 
	\end{multline}
\end{subequations}

Let us now consider the evolution of the QKR when driven with a Fibonacci sequence of kicks. For $\tau\ll 1$ such that Eq.~\eqref{eq_fib_approx} is satisfied, the evolution operator assumes the form,
\begin{equation}\label{eq_U_fibo}
	U(N,0) = \cdots e^{-iL_2}e^{-iL_1}e^{-iL_2}e^{-iL_1}e^{-iL_1}e^{-iL_2}e^{-iL_1}
\end{equation}
Our purpose is to derive an approximate expression for the evolution operator by progressively approximating the unitary operator for adjacent time intervals. Let us denote the evolution over the first two time intervals: $U(2,0)\approx e^{-iL_{2}}e^{-iL_1}=e^{-i\Theta_{12}}$, where $\Theta_{12}$ is to be computed using the BCH formula. As before, we calculate the commutators,
\begin{subequations}
	\begin{equation}
		\left[-iL_{2}, -iL_1\right]= i\left(\mathcal{K}_{2}-\mathcal{K}_{1}\right)\frac{\tau}{2}\Big(\hat{l}\sin\hat{\theta}+\sin\hat{\theta}\hat{l}\Big)+\mathcal{O}(\tau^2),
	\end{equation}
	\begin{equation}
		\left[-iL_1,\left[-iL_1,-iL_{2}\right]\right]=i\mathcal{K}_1\left(\mathcal{K}_{2}-\mathcal{K}_1\right)\tau\sin^2\hat{\theta}+\mathcal{O}(\tau^2),
	\end{equation}
	\begin{equation}
		\left[-iL_{2},\left[-iL_{2},-iL_{1}\right]\right]=-i\mathcal{K}_{2}\left(\mathcal{K}_{2}-\mathcal{K}_1\right)\tau\sin^2\hat{\theta}+\mathcal{O}(\tau^2).
	\end{equation}
\end{subequations}
It can be straightforwardly verified from the above expressions that all other higher order commutators, such as $\left[L_1,\left[L_1,\left[L_1,L_{2}\right]\right]\right]$, will contribute terms which are at least quadratic in order $\tau$. Retaining terms up to linear order in $\tau$, we find,
\begin{widetext}
\begin{equation}
	-i\Theta_{12} = -i\beta(2)L_2-i\alpha(2)L_1+\delta(2)\left[-iL_2,-iL_1\right]\non\\+\eta_1(2)\left[-iL_1,\left[-iL_1,-iL_2\right]\right]+\eta_2(2)\left[-iL_2,\left[-iL_2,-iL_1\right]\right],
\end{equation}
\end{widetext}
where $\alpha(2)=\beta(2)=1$, $\delta(2)=1/2$ and $\eta_1(2)=\eta_2(2)=1/12$. \\

We can now build the unitary operator as follows. After three kicks, the evolution operator can be approximated as $U(3,0)=e^{-iL_1}e^{-i\Theta_{12}}= e^{-i\Theta_{13}}$, where $\Theta_{13}$ is found to be,
\begin{multline}\label{eq_L13}
	-i\Theta_{13} = -2iL_1 - iL_2 -\frac{1}{6}\left[-iL_1,\left[-iL_1,-i L_2\right]\right]\\+\frac{1}{6}\left[-iL_2,\left[-iL_2,-iL_1\right]\right].
\end{multline}
A careful inspection reveals that when the $N^{th}$ kick is $\mathcal{K}_1$, the expansion coefficients obey the following recursion relations,
\begin{subequations}\label{eq_coeff_k1}
	\begin{equation}
		\delta(N) = \delta(N-1) -\frac{1}{2}\beta(N-1)
	\end{equation}
	\begin{equation}
		\eta_1(N) = \eta_1(N-1) - \frac{1}{2}\delta(N-1)+\frac{1}{12}\beta(N-1)\Big(1-\alpha(N-1)\Big)
	\end{equation}
	\begin{equation}
		\eta_2(N) = \eta_2(N-1) + \frac{1}{12}\beta^2(N-1)
	\end{equation}
\end{subequations}
Conversely, when the $N^{th}$ kick is $\mathcal{K}_2$, the recursion relations are given by,
\begin{subequations}\label{eq_coeff_k2}
	\begin{equation}
		\delta(N) = \delta(N-1) +\frac{1}{2}\alpha(N-1)
	\end{equation}
	\begin{equation}
		\eta_1(N) = \eta_1(N-1) + \frac{1}{12}\alpha^2(N-1)
	\end{equation}
	\begin{equation}
		\eta_2(N) = \eta_2(N-1)+ \frac{1}{2}\delta(N-1)+\frac{1}{12}\alpha(N-1)\Big(1-\beta(N-1)\Big)
	\end{equation}
\end{subequations}

To verify the above recursion relations, we explicitly calculate $\Theta_{14}$ and $\Theta_{15}$ as in Eq.~\eqref{eq_L13}, 
\begin{multline}\label{eq_L14}
	-i\Theta_{14} = -3iL_1 - iL_2 -\frac{1}{2}\left[-iL_2,-iL_1\right]\\-\frac{1}{4}\left[-iL_1,\left[-iL_1, -iL_2\right]\right]+\frac{1}{4}\left[-iL_2,\left[-iL_2,-iL_1\right]\right],
\end{multline}

\begin{eqnarray}\label{eq_L15}
\nonumber-i\Theta_{15} = -3iL_1 - 2iL_2 + \left[-iL_2,-iL_1\right]\\
+\frac{1}{2}\left[-iL_1,\left[-iL_1, -iL_2\right]\right].
\end{eqnarray}                             
We recall from Eq.~\eqref{eq_U_fibo} that the fourth and fifth kicks in the Fibonacci sequence are $\mathcal{K}_1$ and $\mathcal{K}_2$, respectively. Thus, the expansion coefficients in $\Theta_{14}$ and $\Theta_{15}$ satisfy the recursion relations given in Eqs.~\eqref{eq_coeff_k1} and ~\eqref{eq_coeff_k2}, respectively.\\

The recursion relations in Eqs.~\eqref{eq_coeff_k1} and~\eqref{eq_coeff_k2} can be unified using the generating function of the binary Fibonacci sequence, defined as,
\begin{equation}
	\gamma(N)=\lfloor (N+1)G\rfloor - \lfloor NG\rfloor,
\end{equation} 
where $G=(\sqrt{5}+1)/2$ is the golden ratio and $\lfloor x\rfloor$ denotes the greatest integer less than or equal to $x$. For any positive integer $N$, $\gamma(N)\in\{1,2\}$. The function $\gamma(N)-1$ is therefore a Boolean function and it generates the required Fibonacci sequence. We show this below by explicitly evaluating it for $N=1,2,3, \dots13$,
\begin{center}
	\begin{tabular}{|c|c|c| c | c | c | c | c | c | c | c | c | c | c |}
		\hline
		$N$&~1&~2&~3&~4&~5&~6&~7&~8&~9&10&11&12&13 \\\hline
		$\gamma(N)-1$&1&0&1&1&0&1&0&1&1&0&1&1&0\\\hline
	\end{tabular}
\end{center} 
Substituting $\mathcal{K}_1$ and $\mathcal{K}_2$ in place of $1$ and $0$ in the second row of the table above, we recover the Fibonacci sequence defined in Eq.~\eqref{eq_fibo_seq} of Sec.~\ref{sec_model}.\\

Using the generating function $\gamma(N)$ defined above, the coefficients $\beta(N)$ and $\alpha(N)$ are immediately evaluated as,
\begin{subequations}
	\begin{equation}
		\beta(N) = \sum_{n=1}^N\Big(2-\gamma(n)\Big),
	\end{equation}
	\begin{equation}
		\alpha(N) = \sum_{n=1}^N\Big(\gamma(n)-1\Big)=N-\beta(N).
	\end{equation}
\end{subequations} 
Having evaluated $\alpha(N)$ and $\beta(N)$, the recursion relation for $\delta(N)$ can be simplified to,
\begin{equation}
	\delta(N) = \sum_{n=1}^N\left[\left(2-\gamma(n)\right)\frac{\alpha(n-1)}{2}-\left(\gamma(n)-1\right)\frac{\beta(n-1)}{2}\right],
\end{equation}
where $\alpha(0)=\beta(0)=0$. Further, it can be verified that if $\gamma(n)=2$, then $\beta(n-1)=\lfloor n/(1+G)\rfloor$. Similarly, if $\gamma(n)=1$, then $\alpha(n-1)=\lfloor nG/(1+G)\rfloor$. Substituting in the above expression, we therefore find,
\begin{equation}
	\delta(N)=-\frac{1}{2}\sum_{n=1}^N\left[\Big(\gamma(n)-1\Big)\Big(n-1\Big)-\Bigl\lfloor\frac{nG}{1+G}\Bigr\rfloor\right],
\end{equation}
where we have used the relation $\lfloor nG/(1+G)\rfloor+\lfloor n/(1+G)\rfloor=n-1$. Finally, the coefficients $\eta_1(N)$ and $\eta_2(N)$ can be evaluated as,
\begin{widetext}
\begin{equation}
	\eta_1(N)=\frac{1}{12}\sum_{m=1}^N\Big[\left(2-\gamma(n)\right)\Bigl\lfloor \frac{nG}{1+G}\Bigr\rfloor^2\non\\+\left(1-\gamma(n)\right)\Big(6\delta(n-1)-(2-n)\Bigl\lfloor \frac{n}{1+G}\Bigr\rfloor-\Bigl\lfloor \frac{n}{1+G}\Bigr\rfloor^2\Big)\Big],
\end{equation}
\begin{equation}
	\eta_2(N)=\frac{1}{12}\sum_{m=1}^N\Big[\left(\gamma(n)-1\right)\Bigl\lfloor \frac{n}{1+G}\Bigr\rfloor^2\non\\+\left(2-\gamma(n)\right)\Big(6\delta(n-1)+(2-n)\Bigl\lfloor \frac{nG}{1+G}\Bigr\rfloor+\Bigl\lfloor \frac{nG}{1+G}\Bigr\rfloor^2\Big)\Big],
\end{equation}
\end{widetext}

\section{Emergence of effective Fibonacci Hamiltonian}\label{app_eff}
As discussed in Sec.~\ref{sec_perturb}, the NECs for terms up to order $\mathcal{O}(\tau)$ either saturate to steady values or oscillate when observed at Fibonacci instants.
Indeed, using the so-called `local deflation rule' for the Fibonacci sequence, one can show that the asymptotic values of the NECs for $\mathcal{N}\gg1$ are given by\cite{andrew18},
\begin{subequations}
	\begin{equation}
		\frac{\alpha(F(\mathcal{N}))}{F(\mathcal{N})}=\frac{1}{G}	
	\end{equation}
	\begin{equation}
		\frac{\beta(F(\mathcal{N}))}{F(\mathcal{N})}=\frac{1}{G^2}	
	\end{equation}
	\begin{equation}
		\frac{\delta(F(\mathcal{N}))}{F(\mathcal{N})}=-\frac{1}{G^3}	
	\end{equation}
	\begin{equation}
		\frac{\eta_1(F(\mathcal{N}))}{F(\mathcal{N})}=\frac{1}{12}\left[\frac{1}{G^4}+(-1)^\mathcal{N}\left(\frac{2}{G}+\frac{1}{G^2}\right)\right]	
	\end{equation}
	\begin{equation}
		\frac{\eta_2(F(\mathcal{N}))}{F(\mathcal{N})}=\frac{1}{12}\left[	\frac{1}{G^5}-(-1)^\mathcal{N}\left(\frac{2}{G^2}+\frac{1}{G^3}\right) +\frac{1}{G^2}\right]	
	\end{equation}
\end{subequations}

%

We shall henceforth denote the saturation values of the NECs at Fibonacci instants as $\bar{\alpha}$, $\bar{\beta}$, $\bar{\delta}$, $\bar{\eta}_1$, $\bar{\eta}_2$,  where $\bar{\eta}_1$ and $\bar{\eta}_2$ correspond to the mean of the oscillating values of $\eta_1(N)/N$ and $\eta_2(N)/N$, respectively. Substituting the saturation values derived above , the {propagator} at Fibonacci instants $U_{F(\mathcal{N})}$ can be expressed in terms of an \textit{effective Fibonacci propagator} $U_{fi}$, such that,
\begin{subequations}\label{eq_eff_prop}
	\begin{equation}
		U_{N=F(\mathcal{N})}\approx U_{fi}^{F(\mathcal{N})}= e^{-iH_{fi}F(\mathcal{N})}
	\end{equation}
	where,
	\begin{eqnarray}\label{eq_eff_fib_ham}
		\nonumber H_{fi}=\bar{\alpha}L_1+\bar{\beta}L_2+\bar{\delta}[L_2,L_1]\\
		+\bar{\eta}_1[L_1,[L_1,L_2]]+\bar{\eta}_2[L_2,&[&L_2,L_1]],
	\end{eqnarray}
\end{subequations}
is defined as the \textit{effective Fibonacci Hamiltonian}.

\bibliography{reference}

\begin{thebibliography}{42}
\bibitem [{3d_()}]{3d_rotor_note}%
  \BibitemOpen
  \href@noop {} {}\bibinfo {note} {The kinetic energy terms in the Hamiltonian
  of the three dimensional rotor are linear in momentum in two of the
  orthogonal directions}\BibitemShut {NoStop}%
\bibitem [{hs_()}]{hs_equivalence}%
  \BibitemOpen
  \href@noop {} {}\bibinfo {note} {An alternate common way of representing the
  evolution operator is, $U_N =
  \exp\left[-i\frac{\hat{p}^2}{2\hbar_s}\right]\exp\left[-i\frac{\tilde{K}_N}{\hbar_s}\cos\hat{\theta}\right]$
  where $\hbar_s=\hbar T/I$ is the effective Planck's constant,
  $\hat{p}=\hbar_s\hat{l}$ and $\tilde{K}_N=K_N\hbar_s$. We work with natural
  units $\hbar=1$ and set $T/I=\tau$, which implies that $\tau$ can
  equivalently considered as the effective Planck's constant.}\BibitemShut
  {Stop}%
\end{thebibliography}

\end{document}